%% file: main_v2.tex
\documentclass[a4paper,11pt]{article}
\pdfoutput=1 

\usepackage{jcappub} 
\usepackage{xspace}
\usepackage{aas_macros}
\usepackage[T1]{fontenc} 
\usepackage[dvipsnames,table]{xcolor}
\usepackage{scalerel}
\usepackage{tikz}
\usetikzlibrary{svg.path} 
\usepackage{cleveref} 
\usepackage{booktabs}

\usepackage{orcidlink}

\title{\boldmath  DESI 2024: Constraints on Physics-Focused Aspects of Dark Energy using DESI DR1 BAO Data}

\include{defs}
\input{DESI-2024-0420_author_list}

\abstract{Baryon acoustic oscillation data from the first year of the Dark Energy Spectroscopic Instrument (DESI) provide near percent-level precision of cosmic distances in seven bins over the redshift range $z=0.1$--$4.2$. We use this data, together with other distance probes, to constrain the cosmic expansion history using some well-motivated physical classes of dark energy. In particular, we explore three physics-focused behaviors of dark energy from the equation of state and energy density perspectives: the thawing class (matching many simple quintessence potentials), emergent class (where dark energy comes into being recently, as in phase transition models), and mirage class (where phenomenologically the distance to CMB last scattering is close to that from a cosmological constant $\Lambda$ despite dark energy dyAll three classes fit the data at least as well as $\Lambda$CDM, and indeed can improve on it by $\Delta\chi^2\approx -5$ to $-17$ for the combination of DESI BAO with CMB and supernova data, while having one more parameter. The mirage class does essentially as well as $w_0w_a$CDM while having one less parameter. These classes of dynamical behaviors highlight worthwhile avenues for further exploration into the nature of dark energy. 
} 

\begin{document}

\maketitle
\flushbottom

\section{Introduction}

Cosmic distances and the cosmic expansion rate contain information on the matter and energy contents of the universe. Redshift surveys can measure these at many epochs in cosmic history and so are especially valuable in separating the contributions and studying their evolution. In particular, a central question in cosmology is the nature of cosmic acceleration: does it originate from a cosmological constant $\Lambda$ or a dynamically evolving dark energy? 

Dark energy can be explored through the use of particular models, general parametrizations, or in a model-independent manner. In the absence of a fundamental theory pointing to a compelling model, and for robustness, many analyses take one of the latter two routes. The standard parametrization for the dark energy equation of state $w(a)=w_0+w_a(1-a)$ originated from exact solutions of the dark energy physical dynamics and has been demonstrated to be accurate to $\sim0.1\%$ \cite{Linder:2002et,dePutter:2008wt} in matching distances and expansion rates over a wide array of models. Results for $w_0w_a$CDM cosmology using DESI data are presented in 
\cite{DESI2024.VI.KP7A}. 
Model-independent or at least model-agnostic approaches, e.g.\ Crossing Statistics, Gaussian Processes, values in redshift bins, etc., 
are investigated in \cite{Calderon:2024uwn} and forthcoming works for DESI data. 

Here, we take a middle path, with general parametrizations informed by physics properties. Thus we do not use specific models, but are focused by the physics into classes of dynamical evolution. One perspective is to address the evolution in terms of the dark energy equation of state, specifically the thawing class, and another in terms of the dark energy density, specifically where dark energy arises quickly, whether through a phase transition or a rapid growth ($w\ll -1$). 
A third class is the mirage class \cite{Linder:2007ka}, where an apparent $w=-1$ when forced to a constant value actually hides dynamics. 
These are chosen because previous data pointed to these physical properties as compatible with observations, at least as well as the cosmological constant, and they describe broad classes of characteristic behavior, rather than a single model. 

New, highly precise data have become available from Data Release 1 of the DESI baryon acoustic oscillations (DESI BAO) measurements. The data are discussed in detail in \cite{DESI2024.III.KP4,DESI2024.IV.KP6,DESI2024.VI.KP7A}. Here we investigate their implications for the dark energy physics, using the physics-focused classes above. \Cref{Sec:Theory} describes the 
physics-focused classes for the dark energy equation of state and density. 
In \cref{sec:data} we review the DESI distance data in seven redshift ranges, as well as other data sets used in combination. 
Constraints on cosmology and the physics implications are discussed in \cref{Sec:Results}, with conclusions in \cref{Conclusion}.

\section{Physics-focused classes}\label{Sec:Theory}
The current data, including DESI BAO, favors dynamical dark energy over a simple cosmological constant at various levels of significance when different data set combinations are used, as shown in \cite{DESI2024.VI.KP7A}. This motivates consideration of a wide variety of dark energy behaviors. 

However, there is no compelling physics-based theory for dark energy differing from $\Lambda$, so one tends to adopt a more phenomenological approach. Here, we want to retain physics to a significant extent and use classes of dark energy properties consistent with the data, and that are more general than specific models. 

Dark energy properties enter the measurements through their 
impact on cosmic distances (we do not here consider growth probes of 
large scale structures). This follows from the Friedmann equations, 
and we can write the dark energy influence as (assuming a spatially 
flat universe) 
\begin{equation}\label{eq:H2}
    \frac{H^2(z)}{H_0^2}=\Omega_{\rm m,0}(1+z)^3+\Omega_{\rm r,0}(1+z)^4+(1-\Omega_{\rm m,0}-\Omega_{\rm r,0})\;\fde(z)\ ,
\end{equation} 
where $\Omega_{\rm m,0}$ and $\Omega_{\rm r,0}$ are the present 
fractions of the critical energy density in matter and radiation 
respectively, and $\fde(z)$ describes the 
dark energy density evolution. The dark energy equation of state, 
or pressure to energy density ratio, then comes from the continuity 
equation as 
\begin{equation}\label{eq:derfz}
      w(z) = -1+\frac{1}{3} \frac{d\,\ln \fde(z)}{d\ln{(1+z)}} \ .
\end{equation} 
For example, 
\begin{equation} 
w(a)=w_0+w_a(1-a) \quad\Leftrightarrow \quad \fde(a)=a^{-3(1+w_0+w_a)}\,e^{-3w_a(1-a)}\ , 
\end{equation} 
where the scale factor $a=1/(1+z)$. 
The key then is seeing how physics informs the dark energy 
density $\fde(z)$ or equation of state $w(a)$. 
We assume throughout that the dark energy fluid sound speed $c_s^2=1$.

\subsection{Dark Energy Equation of State: Thawing Physics} \label{sec:thaw} 

While writing the dark energy equation of state as $w(a)=w_0+w_a(1-a)$ 
\cite{2001IJMPD..10..213C,Linder:2002et} 
has been shown to be highly accurate for a wide variety of models \citep{Linder:2002et,dePutter:2008wt}, the physics does not actually predict that any 
{\it arbitrary\/} combination of $w_0$ and $w_a$ is equally valid. Basic 
physics -- evolution of the dark energy field through the long history 
of radiation and matter domination in the presence of Hubble friction -- 
calls out two regions of the phase space as preferred, known as the 
thawing and freezing regions \cite{Caldwell:2005tm}. Other regions of the phase 
space arise only due to extraordinary circumstances, e.g.\ fine tuning, 
noncanonical kinetic structure, or nongravitational interactions. 
The freezing region, with $w_a>0$, tends not to be compatible with 
observations; indeed DESI BAO plus other probes disfavors it at 
$\sim3\sigma$ \cite{DESI2024.VI.KP7A}. Thus we focus on exploring the thawing physics class. 

Thawing physics arises because, during the long cosmic history, the 
Hubble friction was high enough to overcome dark energy dynamics, 
causing it to act like a cosmological constant. Only recently, as 
the Hubble expansion rate declined sufficiently, was dark energy 
released to allow dynamics (``thawed''). This describes a broad 
variety of particle physics models for dark energy, including 
pseudo-Nambu-Goldstone bosons (PNGB \cite{Frieman:1995pm}; e.g.\ axions), the linear 
potential \cite{linde86} with its shift symmetry shielding against 
quantum corrections, and many monomial potentials (e.g.\ the 
standard quadratic $V\sim m^2\phi^2$ and quartic $V\sim\lambda\phi^4$).

One of the great virtues of the $w_0$--$w_a$ parametrization 
is that it acts as a calibration relation for the physics. Not 
only are the thawing fields in the same class, but $w_0$--$w_a$ 
calibrates their evolution $w(a)$, bringing their phase space tracks 
in $w$--$w'$ into a universal relation \cite{dePutter:2008wt}: 
\begin{equation}\label{eq:thaw-scale}
    w_a\approx -1.58(1+w_0)\ .
\end{equation} 
(The coefficient $-1.58$ comes from fits to the dynamics in \cite{dePutter:2008wt}, e.g.\ see Eq.~(1) of \cite{Linder:2015zxa}.)

Another approach to thawing dynamics is to account for the Hubble 
friction freeze in the past plus an algebraic factor describing 
the thawing, roughly related to the ratio of the frozen dark energy 
density to the matter density, $\sim a^3$. Again, these are general 
characteristics of the thawing class as a whole, and so not model 
dependent in the usual sense. Following \cite{Linder:2007wa,Linder:2015zxa} we have 
\begin{equation}\label{eq:thaw-flow}
    1+w(a)=(1+w_0)\,a^3\left(\frac{3}{1+2a^3}\right)^{2/3}\ .
\end{equation}

Note that both the calibration and algebraic forms have simply one 
parameter more than $\Lambda$CDM. They have also both been demonstrated 
to have accuracy better than 0.1\% in matching distances $d(z)$ and 
Hubble expansion rates $H(z)$ of the exact physics \cite{Linder:2015zxa}. 
Describing the thawing class by either the calibration or algebraic forms gives virtually 
identical results (see \cref{fig:flow-vs-scale} in Appendix~\ref{apx:calalg}), adding support for the model-independent nature of the analysis.

\subsection{Dark Energy Density: Emergent Physics} \label{sec:gede}

In contrast to the previous subsection, we now consider the dark 
energy density, rather than the equation of state, and a rapid emergence 
or transition rather than a slow thaw. In this class of physical behavior, 
dark energy is negligible (or vanishes) above moderate redshift 
(say $z\approx2$) but its energy density quickly grows at lower 
redshift (implying $w<-1$) before leveling off to a constant in 
the future (and so $w\to -1$). Physical examples of this behavior 
include phase transitions such as vacuum metamorphosis 
\cite{Parker:2000pr,Caldwell:2005xb} and dark energy as a critical phenomena~\cite{Banihashemi:2018has,Banihashemi:2020wtb}. The density of dark energy as a critical phenomenon can behave similarly to the magnetization of the Ising model and effectively emerges at a particular time (redshift) corresponding to the critical temperature in the model~\cite{Banihashemi:2020wtb}.  

Phenomenological Emergent Dark Energy (PEDE) model~\cite{Li:2019yem,Pan:2019hac} has been introduced as a zero freedom dark energy model where dark energy has no effective presence in the past and effectively emerges in the late Universe. The model was generalised as Generalised Emergent Dark Energy (GEDE)~\citep{Li:2020ybr}, to include both PEDE and $\Lambda$ as two limits of the parametric form and to include a larger class of emergent dark energy behaviors.  \\

The evolution of the energy density in GEDE is given by \citep{Li:2020ybr,Yang:2021eud} 
\begin{equation}\label{eq:GEDE}
    \fde(z)=\frac{1-\tanh{\left(\Delta\times\log_{10}(\frac{1+z}{1+z_t})\right)}}{1+\tanh{\left(\Delta\times\log_{10}(1+z_t)\right)}}~,
\end{equation} 
where $\Delta$ is a free parameter, determining the steepness of the transition, and $z_t$ is a derived quantity determined by solving $\rho_m(z_t)=\rho_{\rm DE}(z_t)$.
The corresponding equation of state for GEDE is 
\begin{equation}
    w(z)=-1-\frac{\Delta}{3\ln{(10)}}\left[1+\tanh\left(\Delta\log_{10}\left(\frac{1+z}{1+z_t}\right)\right)\right]~.
\end{equation}
Note that $\fde(z)$ goes from much less than one for $z\gg z_t$ to one today to a finite value greater than one in the future (de Sitter state), while $w(z)$ goes from $-1-2\Delta/(3\ln 10)$ at $z\gg z_t$ to $-1$ in the future.

\subsection{Dark Energy: Mirage Physics} \label{sec:mirage} 

Another, more phenomenological class is that of mirage 
dark energy \cite{Linder:2007ka}. This originated as a way to match the 
CMB distance to last scattering from \lcdm\ but for some evolving dark energy equation of state $w(a)$. More generally, it 
will appear to yield a constant $w=-1$ 
for data combinations with a pivot point, or greatest 
sensitivity to dark energy equation of state, 
around $a\approx0.7$. The condition becomes 
\begin{equation}
    w_a=-3.66\,(1+w_0)\ . \label{eq:mirage} 
\end{equation} 
(The coefficient $-3.66$ comes from Eqs.~(1) and (3) of \cite{Linder:2007ka} and varies by a couple percent over the range  $\Omo\in [0.25,0.35]$.)

Interestingly, DESI BAO DR1 gives a confidence 
contour in the $w_0$--$w_a$ plane that follows this 
closely, and indeed delivers a $w\approx-1$ fit when assuming $w=\,$const (e.g.\ see Fig.~5 of \cite{DESI2024.VI.KP7A}). We emphasize, however, as demonstrated in this 
article and \cite{Calderon:2024uwn}, that this does not 
actually mean that $w=\,$const. That may merely be a 
mirage, even for quite rapidly evolving $w(a)$. 
Note furthermore that 
since mirage models match the CMB distance, they 
will also generally closely match the growth of structure (within general 
relativity), as the mirage holds for this as well 
(see \cite{Francis:2007qa} and Fig.~6 of \cite{Linder:2007ka} for demonstration).

As to the physical mechanism behind the mirage, this 
is less clear. Such a crossing of $w(a)=-1$ is a hallmark 
of perhaps a combination of multiple scalar fields, interactions, 
or modified gravity generally involving noncanonical 
kinetic terms and possibly braiding of the scalar 
and tensor degrees of freedom. Rapid emergence of the 
dark energy density (i.e.\ strongly negative $w_a$ and hence $w(a)\ll-1$ at early redshifts) however is not a generic characteristic of 
such mechanisms, and if taken at face value could point 
more to a phase transition mechanism. Many of those, 
however, such as vacuum metamorphosis \cite{Parker:2000pr}, tend toward a de Sitter ($w=-1$) state not $w_0>-1$. It is not clear what reasonable physics would contain both a rapid emergence in density and a crossing of $w(a)=-1$. 
(Speculatively, one could imagine the scalar field responsible for the phase transition having a negative potential or some interaction that would cause crossing of $w(a)=-1$.)

\section{Data and Methodology} \label{sec:data} 

\begin{table}[t]
    \caption{
    Parameters and priors used in the analysis with our modified version of the Boltzmann solver \classy. All of the priors are uniform in the ranges specified below. 
    }
    \label{tab:priors}
\,\\ 
    \centering
    \begin{tabular}{c|c|c}
    \hline
     & parameter & prior/value\\  
    \hline 
    \textbf{background-only} & $\Omo$ &  $\mathcal{U}[0.01, 0.99]$\\
     & $ H_0\rd \; [\kms]$ &  $\mathcal{U}[1000, 100000]$  \\
    \hline 
    \textbf{CMB} & $\omega_{\rm cdm}\equiv\Omega_{\rm cdm}h^2$ & $\mathcal{U}[0.001, 0.99]$ \\
    & $\omega_{\rm b}\equiv\Omega_{\rm b}h^2$ &  $\mathcal{U}[0.005, 0.1]$ \\
    & $\ln(10^{10} A_{s})$ &  $\mathcal{U}[1.61, 3.91]$ \\
    & $n_{s}$ &  $\mathcal{U}[0.8, 1.2]$ \\
    & $H_{0} \; [\kmsMpc]$ &  $\mathcal{U}[20, 100]$  \\
    & $\tau$  &  $\mathcal{U}[0.01, 0.8]$  \\
    \hline \textbf{Thawing/Mirage} & $w_0$ & $\mathcal{U}[-3, 1]$\\
    \hline \textbf{GEDE} & $\Delta$ & $\mathcal{U}[-3, 10]$\\
    \hline $\boldsymbol{w_0 w_a}$ & $w_0$ & $\mathcal{U}[-3, 1]$\\
        & $w_a$ & $\mathcal{U}[-3, 2]$\\
    \hline
    \end{tabular}
\end{table}

The cosmological probes and specific data sets used in our analysis are: 

\begin{itemize}
    \item \textbf{Baryon Acoustic Oscillations (BAO):} 
    We use the compilation of compressed distance quantities $\DMrd$, $\DHrd$, and $\DVrd$ from the first year data release of the Dark Energy Spectroscopic Instrument \cite{DESI2016a.Science,DESI2022.KP1.Instr,DESI2023a.KP1.SV,DESI2023b.KP1.EDR,DESI2024.I.DR1}, where $D_M$ is the transverse comoving distance, $D_H$ the Hubble distance, $D_V$ their isotropic average, and $r_d$ is the sound horizon at the baryon drag epoch. This dataset, abbreviated as ``DESI BAO'', spans seven redshift bins from $z=0.3$ to $z=2.33$ \cite{DESI2024.III.KP4}. 
    We refer the reader to \cite{DESI2024.I.DR1,DESI2024.II.KP3,DESI2024.III.KP4,DESI2024.V.KP5,DESI2024.IV.KP6,DESI2024.VI.KP7A,DESI2024.VII.KP7B} for further details.
     
    \item \textbf{Supernovae Ia (SNe~Ia):} We combine with supernova data from  
    three sets, one at a time: ``PantheonPlus'', a compilation of 1550 supernovae spanning a redshift range from $0.01$ to $2.26$ \cite{Brout:2022vxf}, ``Union3'', containing 2087 SNe~Ia processed through the Unity 1.5 pipeline based on Bayesian Hierarchical Modelling \cite{Rubin:2023ovl}, and ``DES-SN5YR'', a compilation of 194 low-redshift SNe~Ia ($0.025<z<0.1$) and 1635 photometrically classified SNe~Ia covering the range $0.1<z<1.3$ \cite{DES:2024tys}. 
    
    \item \textbf{Cosmic Microwave Background (CMB):} We also include temperature and polarization measurements of the CMB from the Planck satellite \cite{Planck:2018vyg}. In particular, we use the high-$\ell$ TTTEEE likelihood (\texttt{planck\_2018\_highl\_plik.TTTEEE}), together with low-$\ell$ TT (\texttt{planck\_2018\_lowl.TT}) and low-$\ell$ EE (\texttt{planck\_2018\_lowl.EE}) \cite{Aghanim:2019ame}, as implemented in \Cobaya~\cite{Torrado:2020dgo}. Additionally, we include CMB lensing measurements from the combination of NPIPE PR4 from Planck \cite{Carron:2022eyg} and the Atacama Cosmology Telescope (ACT DR6) \citep{madhavacheril2023atacama,ACT:2023dou} using importance sampling. When using the combined Planck+ACT lensing likelihood, we use        \texttt{accurate\_lensing:1} and \texttt{delta\_l\_max:800} options to match CAMB precision settings as recommended by ACT.
\end{itemize}

In our analysis, we utilize Markov Chain Monte Carlo (MCMC) sampling to explore the parameter space using the Metropolis-Hastings algorithm \cite{Lewis:2002ah,Lewis:2013hha} as implemented in \Cobaya ~\cite{Torrado:2020dgo}. To facilitate efficient sampling of the CMB Planck likelihoods, we employ the ``fast-dragging''  scheme \cite{Neal:2005}. We have adopted priors similar to \cite{DESI2024.VI.KP7A}, as presented in Table \ref{tab:priors}, and have modified the Boltzmann solver \classy\ \cite{class_2011arXiv1104.2932L,class_2011JCAP...07..034B} incorporating a generalized equation of state for dark energy for the theoretical prediction of observables. We switched to the \texttt{Recfast} option for recombination as it does not assume anything about the equation of state. We assume three neutrino species with $\sum m_\nu=0.06 ~\rm eV$ and $N_{\rm eff} = 3.044$. For the supernovae likelihoods (PantheonPlus, Union3, and DES-SN5YR), we analytically marginalize over the absolute magnitude $M_B$. 
For clarity of presentation, in the main text figures we use PantheonPlus but list constraints from each supernova set in the tables and show their contours in \Cref{apx:snsets}.

\section{Results and Discussions}\label{Sec:Results}

We present the results for each class in turn, showing the cosmological parameter joint posteriors and the reconstructed dark energy equation of state $w(z)$ and energy density $\fde(z)$ for various combinations of data sets. 

\subsection{Thawing}\label{subsec:thaw}
\begin{figure}[h]
    \centering
    \includegraphics[width=0.85\textwidth]{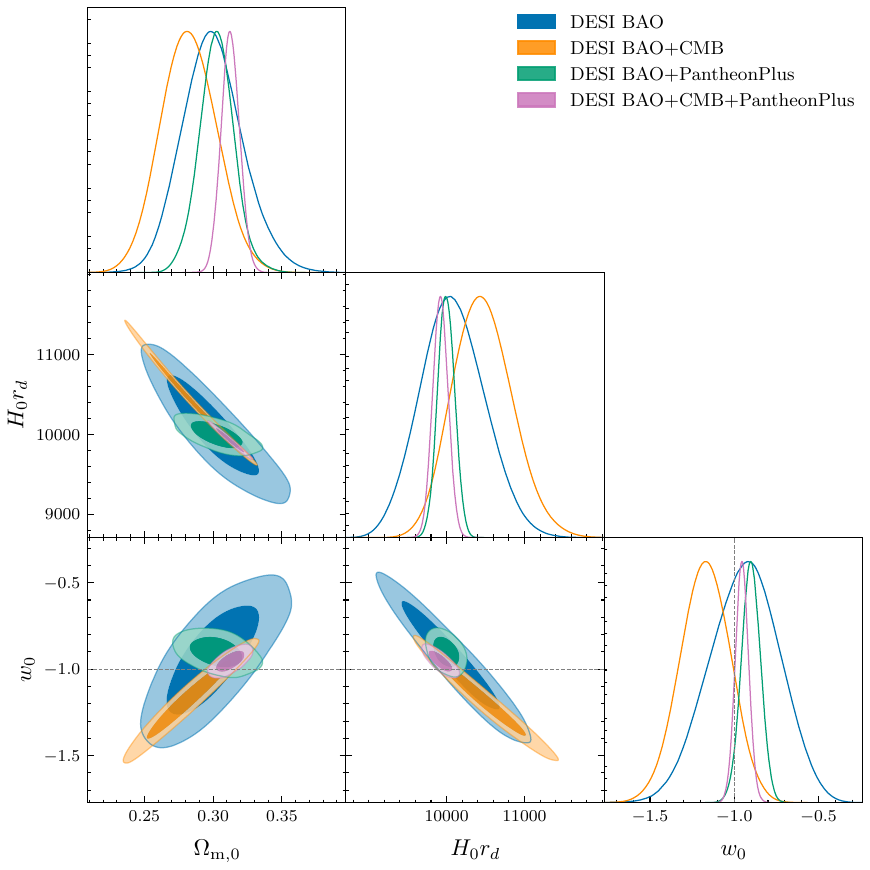}
    \caption{Marginalized constraints within the thawing class of dark energy described by \cref{eq:thaw-scale}, from different combinations of data sets.
    }
    \label{fig:thawing}
\end{figure}
\begin{figure}[h]
    \centering
    \includegraphics[width=\textwidth]{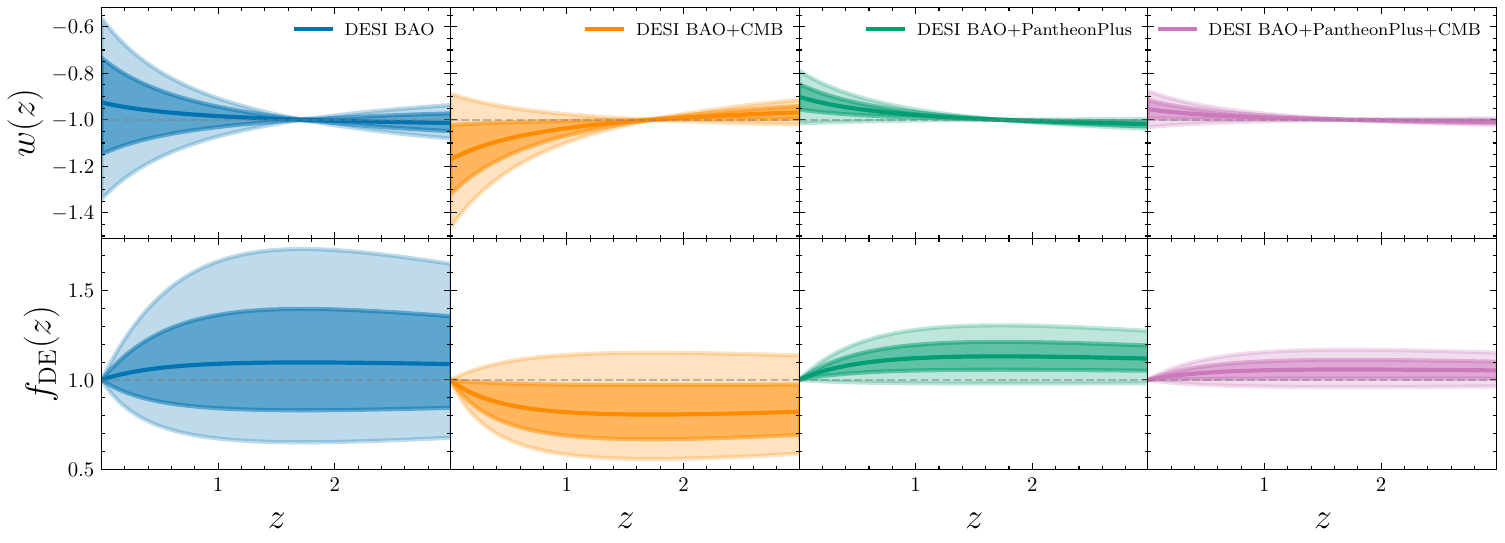}
    \caption{Marginalized constraints on the dark energy equation of state $w(z)$ and energy density $\fde(z)$ for the thawing class, parametrized by \cref{eq:thaw-scale}.  }
    \label{fig:w-thawing}
\end{figure}

\Cref{fig:thawing} shows the joint parameter constraints for the thawing class. The dashed black line corresponds to \lcdm \ ($w_0 = -1$). In the first few rows of \cref{tab:parameter_table2}, we report the marginalized constraints on some of the relevant cosmological parameters and for various data combinations. 
The addition of CMB data to DESI BAO significantly reduces the uncertainty in $w_0$, shifting its value to $< -1$, which also results in $w_a > 0$. However, combining DESI BAO with PantheonPlus yields $w_0 > -1$. A combination of all three datasets provides even tighter constraints, with posteriors peaking at $w_0\gtrsim-1$, hence $w_a < 0$. Using either of the other two supernova datasets instead somewhat strengthens $w_0>-1$. \Cref{fig:w-thawing} illustrates the behaviors for the equation of state $w(z)$ and energy density evolution $\fde(z)$. 

We emphasize that the relations  $w_a(w_0)$ for both the thawing and mirage classes are designed to replicate {\it observations\/}, i.e.\ distances and Hubble parameters to $\sim0.1\%$, not the actual $w(z)$ for a specific model within the class, and hence do not need to have $w(z\gg1)\to-1$, say. 
Indeed $w(z)$ can appear to cross $-1$ even if it actually doesn't, yet still fit the observations exquisitely -- this is well known: 
see Table~1 of \cite{Linder:2015zxa} and Fig.~14 of \cite{Linder:2008pp} for a PNGB model, \cite{Shlivko:2024llw} for a hilltop model, etc. However, for $|w_a|\gtrsim 1$ such unreal crossings of $w(z)=-1$ tend to be in conflict with the CMB distance to last scattering, and such strong crossings tend in fact to be real.

\subsection{Generalized Emergent Dark Energy}
\begin{figure}[h]
    \centering
    \includegraphics[width=0.85\textwidth]{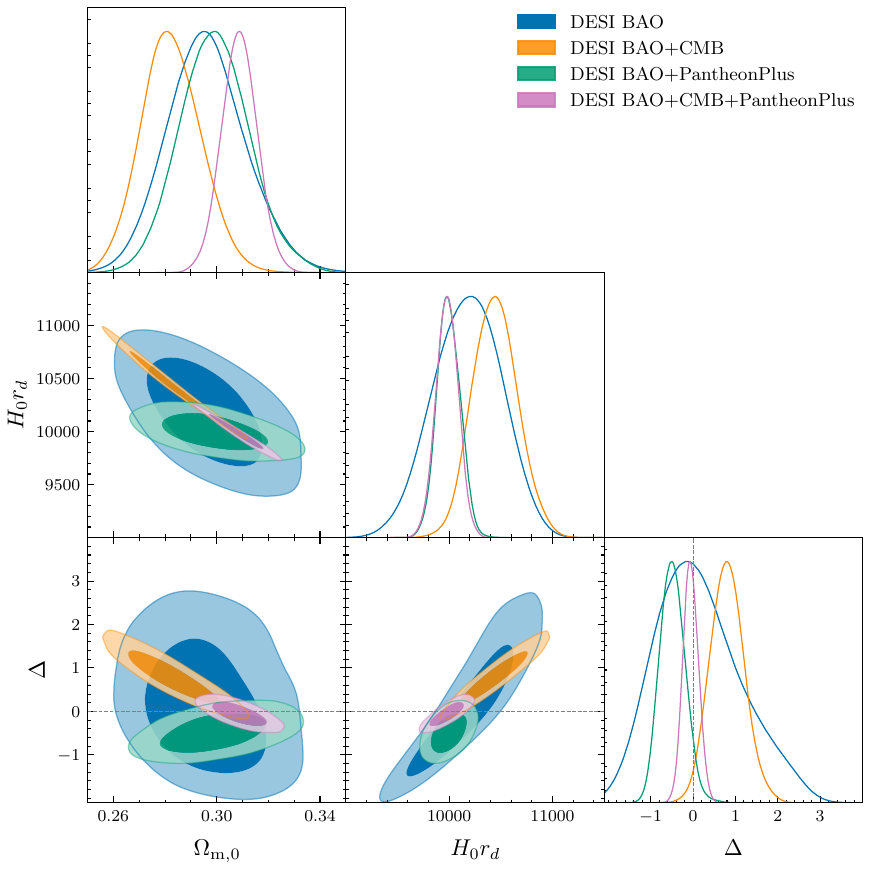}
    \caption{Marginalized constraints within the Generalized Emergent Dark Energy (GEDE) class described by \cref{eq:GEDE}.} 
    \label{fig:GEDE}
\end{figure}

\begin{figure}[h]
    \centering
    \includegraphics[width=\textwidth]{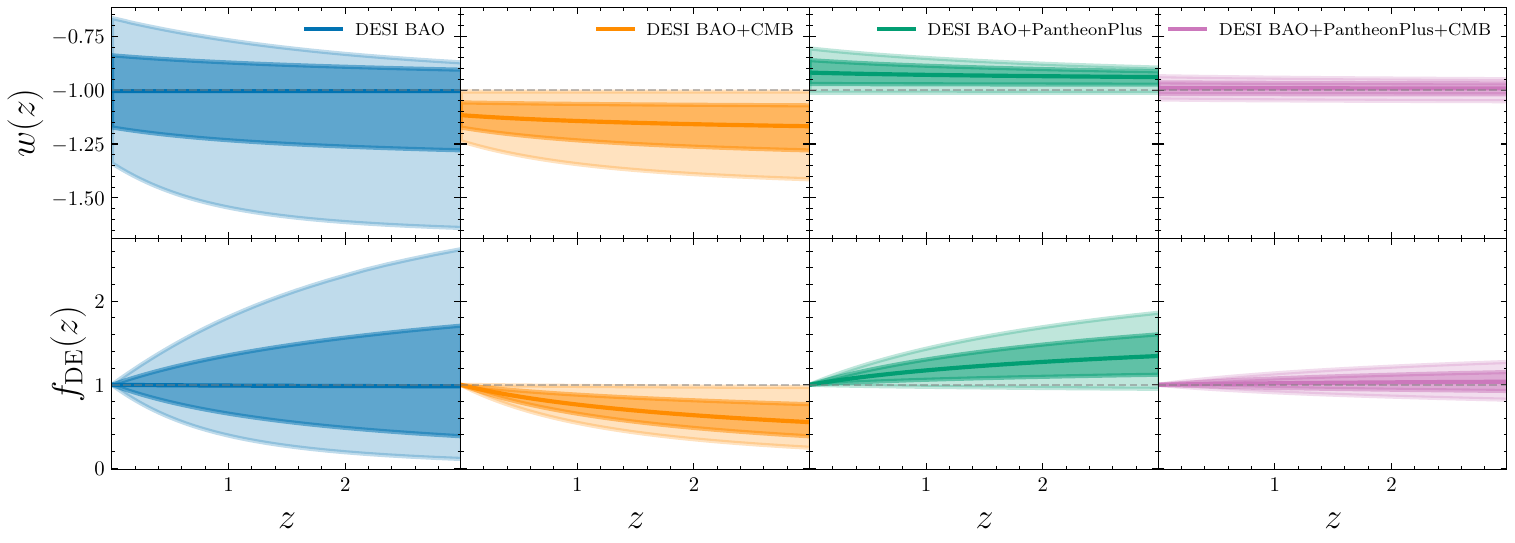}
    \caption{Marginalized constraints on the dark energy equation of state $w(z)$ and energy density $\fde(z)$ for the GEDE model, parametrized by \cref{eq:GEDE}.  }
    \label{fig:w-GEDE}
\end{figure}

The GEDE analysis proceeds similarly, with \cref{fig:GEDE} showing 
its constraints on parameters, using the same datasets. When considering DESI BAO alone, the dataset is broadly consistent with $\Delta = 0$, corresponding to \lcdm. However, when combined with CMB, $\Delta$ peaks at a positive value ($\Delta\simeq1$, corresponding to PEDE \cite{Li:2019yem}), indicating that dark energy emerges at late times. 
Adding PantheonPlus, GEDE prefers a negative value of $\Delta$, which corresponds to the injection of energy at earlier redshifts. However, combining all three datasets results in a peak near $\Delta=0$, indicating that dark energy density remains roughly constant throughout evolution and that GEDE is not preferred over $\Lambda$CDM. 
Note that the model-agnostic reconstructions of dark energy in \cite{Calderon:2024uwn} do seem to indicate a sharp emergence of dark energy in the recent past. The issue is that GEDE 
sharply emerges, but asymptotes to $w=-1$ rather than crossing it. 
Increasing $\Delta$ fits the $z\approx1$ data better than $\Lambda$CDM but the $z\lesssim0.5$ data prefers $w>-1$ and so GEDE is worse than 
$\Lambda$CDM there, resulting in GEDE ``mellowing'' to approach $\Lambda$CDM behavior (and so, as we will see, not having a particularly advantageous goodness of fit). 
If there were an emergent model that also crossed $w=-1$ then this might provide a superior fit to data, but physics motivation for such behavior is not obvious. 
\Cref{fig:w-GEDE} exhibits the uncertainty bands for the reconstructed equation of state $w(z)$ and energy density evolution $\fde(z)$.


\subsection{Mirage}
\begin{figure}[h]
    \centering
    \includegraphics[width=0.85\textwidth]{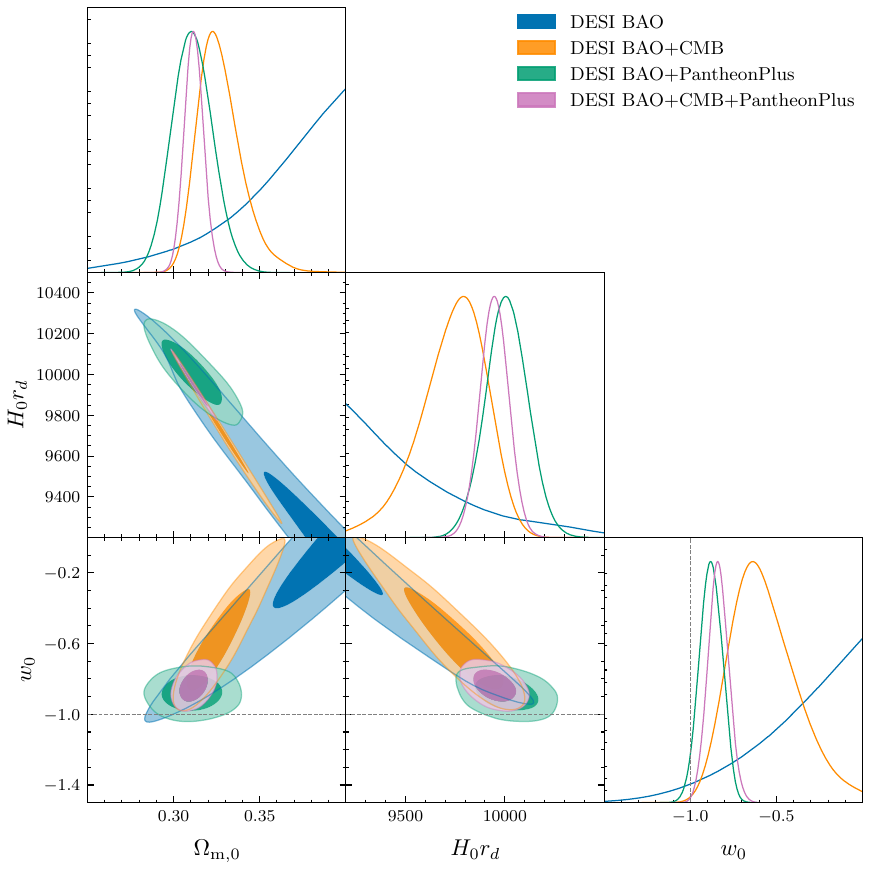}
    \caption{Marginalized constraints within the mirage class of dark energy described by \cref{eq:mirage}. 
    }
    \label{fig:mirage}
\end{figure}

The mirage class has cosmological 
parameter constraints illustrated in \cref{fig:mirage}. 
Again, the preference is pulled off \lcdm\ (which is a member 
of this class, where the mirage is real). 
A best fit of $w_0\approx-0.8$, and hence $w_a\approx-0.7$, is 
quite consistent with DESI BAO data, including in combination 
with other data sets such as CMB and supernovae. 
One can make $w_0$ even less negative (and hence $w_a$ more negative), i.e.\ strengthen 
the mirage, if one compensates by decreasing the late 
time dark energy density (increasing $\Omo$), as seen in 
\cref{fig:mirage}. 

\begin{figure}[h]
    \centering
    \includegraphics[width=\textwidth]{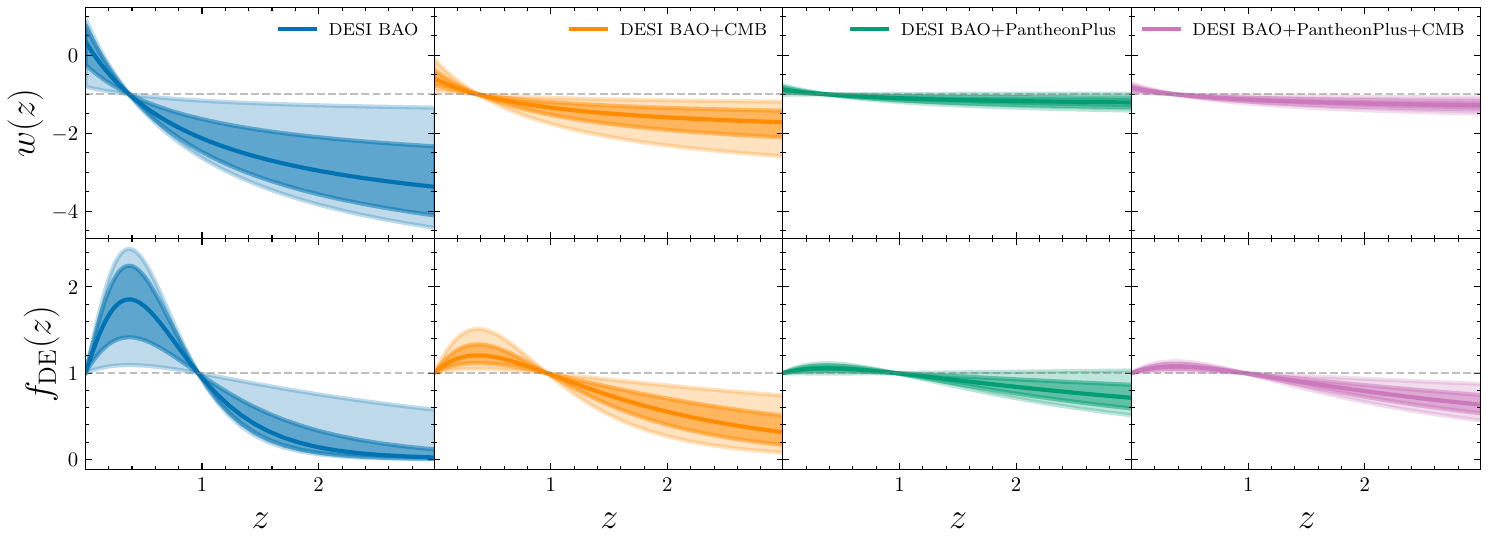}
    \caption{Marginalized constraints on the equation of state $w(z)$ for the mirage class, parametrized by \cref{eq:mirage}.  }
    \label{fig:w-mirage}
\end{figure}

At earlier times, the strongly negative $w_a$ implies a strongly negative $w(a)$, and hence very little dark energy density, 
before rapidly increasing in energy density while 
crossing $w(a)=-1$. This effectively is acting like GEDE at higher 
redshift and the thawing class at lower redshift. 
In \Cref{fig:w-mirage}, we show the reconstructed equation of state $w(z)$, crossing the $w = -1$ threshold near $z \approx 0.4$, along with the corresponding  evolution of energy density $\fde(z)$; these results agree with the DESI results in \cite{Calderon:2024uwn,DESI2024.VI.KP7A}.


\subsection{Comparison and Discussion} 

\Cref{tab:parameter_table2} summarizes the constraints on relevant cosmological parameters.
One general aspect to note is that the clustering amplitude parameter $S_8$ has a rather consistent and reasonable value for all three classes (note that no direct galaxy clustering growth data is used, only BAO). When combining all three cosmological probes, the values of $H_0$ and $\Omm$ are also quite consistent both across classes and regardless of which supernova dataset is included. For the thawing class, $w_0$ is pulled somewhat off $\Lambda$. For GEDE, $\Delta$ is mostly consistent with zero (hence $\Lambda$). The mirage class is where the strongest deviation from $\Lambda$ is seen, and as we discuss next is where the goodness of fit is best as well. Consistency between a class and $\Lambda$CDM should be viewed not through the 1D confidence intervals, however, but through the joint parameter constraints and the $\Delta\chi^2$ quantification.

\begin{table}
\caption{Constraints on the relevant cosmological parameters
    \vspace{0.5em}}
    \label{tab:parameter_table2}
\centering
\small
\resizebox{\textwidth}{!}{\begin{tabular}{cccccc}
\toprule
Model/Dataset & $H_0 [\rm km\,s^{-1}\,Mpc^{-1}]$ & $\Omega_{\mathrm{m},0}$ & $w_0$ & $\Delta$ & $S_8$ \\
\midrule[2pt] 
\textbf{Thawing} & & & & &\\
DESI BAO+CMB & $71.0^{+2.4}_{-2.8}$ & $0.282\pm 0.020$ & $-1.17\pm 0.15$ & -- & $0.812\pm 0.011$ \\
\midrule
DESI BAO+CMB+Union3 & $66.55\pm 0.95$ & $0.3195\pm 0.0094$ & $-0.906\pm 0.060$ & -- & $0.8223\pm 0.0094$ \\
\midrule
 DESI BAO+CMB+DES-SN5YR & $66.49\pm 0.61$ & $0.3200\pm 0.0065$ & $-0.902\pm 0.038$ & -- & $0.8224\pm 0.0091$ \\
\midrule
 DESI BAO+CMB+PantheonPlus & $67.31\pm 0.66$ & $0.3125\pm 0.0067$ & $-0.954\pm 0.040$ & -- & $0.8208\pm 0.0092$ \\
\midrule[2pt] 
\textbf{GEDE} & & & & &\\

DESI BAO+CMB & $71.0\pm 1.5$ & $0.282\pm 0.012$ & -- & $0.81\pm 0.40$ & $0.8156\pm 0.0092$ \\
\midrule
DESI BAO+CMB+Union3 & $67.6^{+1.0}_{-0.93}$ & $0.3096\pm 0.0089$ & -- & $-0.08\pm 0.25$ & $0.8199\pm 0.0092$ \\
\midrule
DESI BAO+CMB+DES-SN5YR & $66.80\pm 0.69$ & $0.3166\pm 0.0066$ & -- & $-0.28\pm 0.17$ & $0.8207\pm 0.0091$ \\
\midrule
DESI BAO+CMB+PantheonPlus & $67.73\pm 0.74$ & $0.3088\pm 0.0070$ & -- & $-0.06\pm 0.18$ & $0.8198\pm 0.0093$ \\
\midrule[2pt] 
\textbf{Mirage} & & & & &\\
DESI BAO+CMB & $66.2^{+1.3}_{-0.79}$ & $0.3271^{+0.0093}_{-0.015}$ & $-0.56^{+0.15}_{-0.23}$ & -- & $0.842\pm 0.013$ \\
\midrule
 DESI BAO+CMB+Union3 & $66.67\pm 0.59$ & $0.3217\pm 0.0072$ & $-0.657^{+0.096}_{-0.11}$ & -- & $0.838\pm 0.010$ \\
\midrule
DESI BAO+CMB+DES-SN5YR & $67.10\pm 0.44$ & $0.3169\pm 0.0056$ & $-0.742\pm 0.066$ & -- & $0.8332\pm 0.0091$ \\
\midrule
 DESI BAO+CMB+PantheonPlus & $67.55\pm 0.43$ & $0.3117\pm 0.0055$ & $-0.837\pm 0.060$ & -- & $0.8272\pm 0.0092$ \\
\midrule
\end{tabular}}
\end{table}

\Cref{tab:best-fit} presents how the three classes compare to each other, relative to  $\Lambda$CDM, and $w_0w_a$CDM in goodness of fit ($\Delta\chi^2$), for various combinations of datasets. Here, we report $\Delta \chi_\mathrm{MAP}^2$, the difference between the maximum a posteriori of the model and the maximum of the posterior fixing $w_{0} = -1$ or $\Delta = 0$ (i.e.\ to the cosmological constant case). Note that the 
three classes have one more parameter than $\Lambda$CDM and one less 
than $w_0w_a$CDM. 
The first general result of interest is that all three classes have better $\chi^2$ than $\Lambda$CDM. They do have one more parameter but in the combination of all three cosmological probes, the improvement is notably more than one. Note that $w_0w_a$CDM has a significantly better $\chi^2$ than $\Lambda$CDM, even taking into account its two more parameters, as discussed in \cite{DESI2024.VI.KP7A}. However, the $\chi^2$ for the physics-focused classes are often close to the $w_0w_a$CDM values, while having one less parameter. This is especially true for the mirage class, while the thawing class and GEDE appear to be less favored. 
As thawing and mirage are subsets of $w_0w_a$CDM, their $\chi^2$ cannot go below that of $w_0w_a$; for the case of fitting DESI BAO data alone, this appears not to hold for the mirage class, but this is due to the limited prior range of $-3<w_a<2$ (also used for  $w_0w_a$ in \cite{DESI2024.VI.KP7A}) -- see the extended degeneracies in \cref{fig:mirage}. 
When combining DESI BAO with other data this influence of the prior no longer matters. 

The promising $\Delta \chi^2_{\text{MAP}}$ for the mirage class led us to conduct additional nested-sampling runs using the \texttt{PolyChord} sampler \cite{Handley:2015fda} to calculate the Bayesian evidence using the \texttt{anesthetic} package \cite{Handley:2019mfs}. We report the Bayes factors of $|\ln B_{21}|=2.8 \ (0.65)$, $4.2 \ (2.4)$, and $6.4 \ (2.8)$ in favor of the mirage class (compared to $w_0w_a$CDM) over \lcdm\ for the DESI+CMB with PantheonPlus, Union3, and DES-SN5YR data combinations, respectively. These findings suggest a moderate preference for the mirage class over \lcdm\ by the PantheonPlus combination and a strong preference by the Union3 and DES-SN5YR on a Jeffreys’ scale \cite{Jeffreys:1939xee, Trotta:2005ar}.

\begin{table}[ht]
    \centering
    \caption{$\Delta \chisq_\mathrm{MAP} \equiv \chisq_{\rm model}-\chisq_{\Lambda\rm CDM}$ values for the different models and data combinations. The minimum $\chisq$ values were obtained using \texttt{iminuit} \cite{iminuit} and \texttt{Py-BOBYQA} \cite{cartis2018improving,Cartis_2021} minimizer. Note that all data combinations include DESI BAO. 
    } 
$\,$\\ 

\label{tab:best-fit}
    \begin{tabular}{c|c|c|c|c} \hline  \hline 
         Data & $\Delta \chi^2_{{\rm Thawing}}$ &$\Delta \chi^2_{{\rm GEDE}}$ &$\Delta \chi^2_{{\rm Mirage}}$ &  $\Delta \chi^2_{w_0w_a}$
         \\ \hline  
         DESI BAO& $-0.2$ & $-0.04$ & $-5.0$ & $-3.8$\\  
         \hline 
         +CMB& $-0.6$ & $-5.7$ & $-7.6$ & $-8.9$  \\ \hline 
         +PantheonPlus& $-3.2$ & $-3.0$ & $-3.5$ & $-3.5$\\
         +Union3& $-6.3$ & $-5.2$ & $-8.7$ & $-8.9$\\ 
         +DES-SN5YR& $-8.8$ & $-7.7$ & $-10.7$ & $-11.1$\\ 
 \hline 
         +CMB+PantheonPlus& $-0.6$ & $-1.7$ & $-9.0$ & $-9.6$\\ 
         +CMB+Union3& $-3.0$ & $-3.2$ & $-15.2$ & $-15.6$\\ 
         +CMB+DES-SN5YR& $-5.0$ & $-4.8$ & $-17.7$ & $-18.3$\\ \hline  \hline 
    \end{tabular}
\end{table}

\section{Conclusion} \label{Conclusion}

Physics-focused classes can give insight into the nature of dynamical dark energy. Using DESI BAO data combined with different state-of-the-art supernovae compilations (PantheonPlus, Union3, DES-SN5YR) and CMB (Planck and ACT) observations, our main result indicates a preference for evolving dark energy rather than a cosmological constant. This behavior can be very well captured by the mirage class, evolving from $w<-1$ and low energy density at $z\gtrsim1$ to $w>-1$ more recently. Note that this also gives a hump in the dark energy density at $z\approx0.3$--0.5, in agreement with our previous model-agnostic findings \cite{Calderon:2024uwn}. The mirage class of dark energy models has a comparable $\Delta \chi^2$ with that of the $w_0w_a\rm CDM$ model, while having one less degree of freedom. 

The mirage class combines the emergence of dark energy density, perhaps indicative of a phase transition, with the recent evolution of $w(a)$ to less negative values than the cosmological constant of the thawing class. With DESI+SNe~Ia, consistently across the three supernova sets, all three classes have better fits than $\Lambda$CDM and come close to $w_0w_a$CDM (which has one more parameter). Neither thawing nor GEDE have a strong advantage for DESI+CMB, however, and the combination of all three cosmology probes gives a clear advantage to the mirage class over the other two (which are still better fits than $\Lambda$CDM). This preference is reflected in the Bayes factor, showing a moderate to strong preference (depending on the SNe~Ia dataset considered) for the Mirage class over \lcdm.
However, the significance of the Bayesian evidences have to be interpreted cautiously \cite{Koo:2021suo,Keeley:2021dmx}. We leave a detailed model-selection analysis for future works.

Other cosmology parameters such as $H_0r_d$, $\Omm$, and $S_8$ remain near $\Lambda$CDM values when using any of the three classes with the full dataset combination. Together with model independent analyses, such physics-focused classes provide important clues to the physical properties we should seek in dark energy models, beyond the ``blank slate'' characterization of $w_0$--$w_a$. 

The dark energy properties indicated by the data -- consistent with \cite{DESI2024.VI.KP7A,Calderon:2024uwn} -- are rapid evolution from $w(a\ll1)\ll -1$ across $w=-1$ to more recent $w(a\approx1)>-1$, and hence emergent dark energy density at modest redshift while at low redshift an energy density bump together with a slowing down of recent cosmic acceleration (see e.g.\ the $q(z)$ reconstruction in \cite{Calderon:2024uwn}). 
These characteristics do not generally exist simultaneously in the usual dark energy models. Phase transition-type behavior does not generally give $w$ evolving away from $\Lambda$ today, and the thawing class evolving away from $\Lambda$ today does not generally give $w<-1$, let alone $w\ll-1$, in the past. 

If the data and its analysis hold, then we are facing a more complicated dark energy sector than generally treated, possibly involving multiple components or involving special nonlinearities in the action (modified gravity or couplings). Fortunately, further data is imminent, with more DESI BAO data, as well as DESI measurements of redshift space distortions and peculiar velocities that can test cosmic growth and gravity.

\acknowledgments
The authors thank Luis Urena-Lopez for his help with the \texttt{polychord} runs.
We acknowledge the use of the Boltzmann solver \classy~\cite{class_2011arXiv1104.2932L,class_2011JCAP...07..034B} for the computation of theoretical observables, 
\Cobaya\ \cite{Torrado:2020dgo} for the sampling and \gd\ \cite{Lewis:2019xzd} for the post-processing of our results. We also acknowledge the use of the standard \texttt{python} libraries for scientific computing, such as \texttt{numpy} \cite{harris2020array}, \texttt{scipy} \cite{2020SciPy-NMeth} and \texttt{matplotlib} \cite{Hunter:2007}.
This work was supported by the
high-performance computing cluster Seondeok at the Korea Astronomy and Space Science Institute. A.S. would like to acknowledge the support by National Research Foundation of Korea 2021M3F7A1082056, 
and the support of the Korea Institute for Advanced
Study (KIAS) grant funded by the government of Korea.

This material is based upon work supported by the U.S. Department of Energy (DOE), Office of Science, Office of High-Energy Physics, under Contract No. DE–AC02–05CH11231, and by the National Energy Research Scientific Computing Center, a DOE Office of Science User Facility under the same contract. Additional support for DESI was provided by the U.S. National Science Foundation (NSF), Division of Astronomical Sciences under Contract No. AST-0950945 to the NSF’s National Optical-Infrared Astronomy Research Laboratory; the Science and Technology Facilities Council of the United Kingdom; the Gordon and Betty Moore Foundation; the Heising-Simons Foundation; the French Alternative Energies and Atomic Energy Commission (CEA); the National Council of Humanities, Science and Technology of Mexico (CONAHCYT); the Ministry of Science and Innovation of Spain (MICINN), and by the DESI Member Institutions: \url{https://www.desi.lbl.gov/collaborating-institutions}. Any opinions, findings, and conclusions or recommendations expressed in this material are those of the author(s) and do not necessarily reflect the views of the U. S. National Science Foundation, the U. S. Department of Energy, or any of the listed funding agencies.

The DESI collaboration is honored to be permitted to conduct scientific research on Iolkam Du’ag (Kitt Peak), a mountain with particular significance to the Tohono O’odham Nation.

\section*{Data Availability}

The data used in this analysis will be made public along the Data Release 1 (details in \url{https://data.desi.lbl.gov/doc/releases/}).

\appendix
\section{Thawing Class: Calibration vs Algebraic Forms} \label{apx:calalg} 

\begin{figure}[h]
    \centering
    \includegraphics[width=0.8\textwidth]{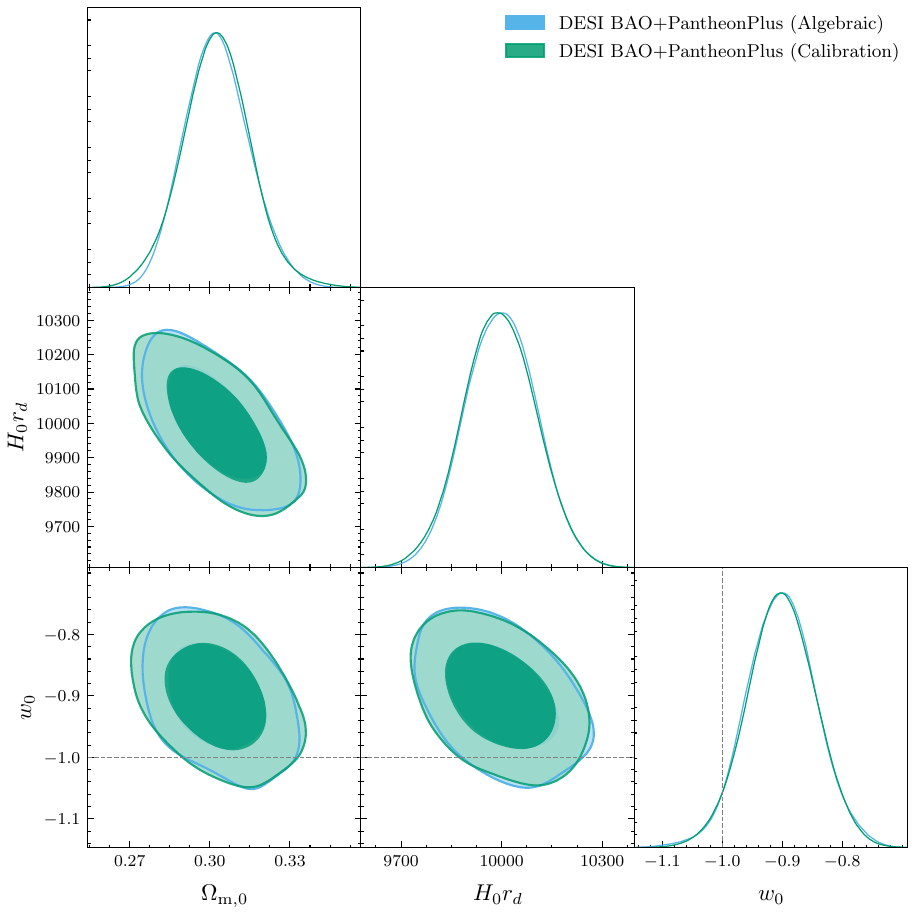}
    \caption{Comparison of the calibration \cref{eq:thaw-scale} vs algebraic \cref{eq:thaw-flow} parametrizations of the thawing class of dark energy, using DESI BAO+PantheonPlus. 
    }
    \label{fig:flow-vs-scale}
\end{figure}

The thawing class encompasses rich physics, including pseudo-Nambu 
Goldstone bosons (PNGB axions), the shift symmetric linear potential, 
and the classic $\phi^2$ and $\phi^4$ potentials. The calibration  
\cref{eq:thaw-scale} and algebraic \cref{eq:thaw-flow} forms were shown to accurately describe 
the exact numerical solutions for observables to $\sim0.1\%$ (e.g.\ 
Table 1 of \cite{Linder:2015zxa}, and \cite{dePutter:2008wt}). 
The former emphasizes the calibration, i.e.\ model independent, aspects of the physics, and the latter the dynamical evolution flow, but each captures the essential physics and \Cref{fig:flow-vs-scale} confirms that the two forms give nearly identical results. 
The specific numbers quoted in this work use the calibration form. 

Note that while the algebraic form \cref{eq:thaw-flow} does not cross $w=-1$ while the calibration form \cref{eq:thaw-scale} does, they both describe the \textit{observations} nearly identically, with excellent accuracy. Thus not every $w(a)$ that crosses $-1$ indicates a physical crossing of $-1$; the forms are designed to describe the observations, not $w(a)$ itself, as emphasized in \cref{subsec:thaw}. However, for $w_0$ too far from $-1$, i.e. $|w_a|$ large enough, such apparent crossings will not fit certain observations like the distance to CMB last scattering, and so $|w_a|\gtrsim 1$ often does point to a real crossing of $-1$ by $w(a)$.

\section{Supernova Data Comparison} \label{apx:snsets} 

\begin{figure}[h]
    \centering
    \includegraphics[width=\textwidth]{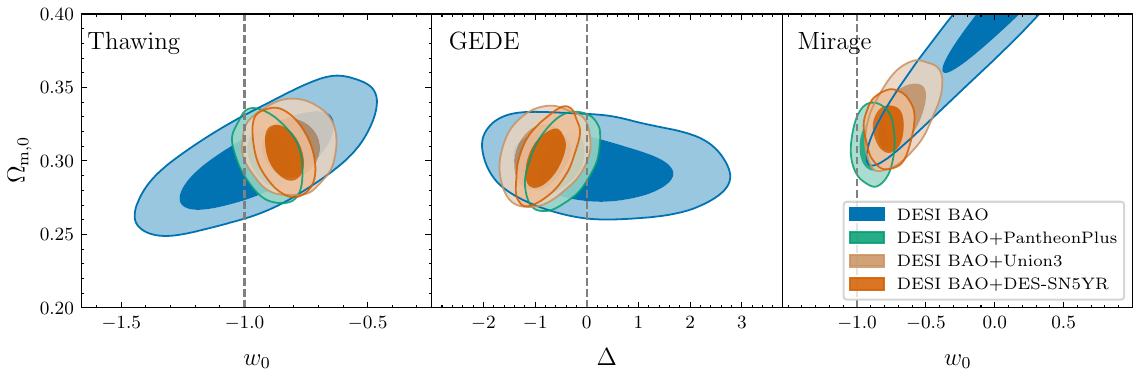}
    \caption{Marginalized posterior distributions using DESI BAO and each supernova dataset. The vertical dashed line indicates $\Lambda$CDM. }
    \label{fig:DESI+SN}
\end{figure}

The figures in the main section of the paper use PantheonPlus as the supernova dataset for clarity of presentation by limiting the number of contours.  \Cref{tab:parameter_table2,tab:parameter_table3} list the parameter constraints for each supernova dataset in turn. Here, 
\Cref{fig:DESI+SN} presents 
the joint confidence contours using DESI BAO in combination with each supernova set in turn. 
The results are quite consistent between each supernova dataset and a similar trend can also be seen in Figure 6 of \cite{DESI2024.VI.KP7A} for \wowacdm. 

\begin{table}[h]
\caption{ Constraints on the relevant cosmological parameters for DESI BAO and DESI BAO+SNe~Ia datasets. 
    \vspace{0.5em}}
    \label{tab:parameter_table3}
\centering
\resizebox{0.95\textwidth}{!}{\begin{tabular}{ccccc}
\toprule
Model/dataset &  $\Omo$ & $H_0r_d$ [$\kms$]  & $w_0$ & $\Delta$ \\
\midrule[2pt] 
\textbf{Thawing} & & & & \\
DESI BAO & $0.300^{+0.020}_{-0.023}$ & $10080\pm 410$ & $-0.94\pm 0.20$ & -- \\
\midrule
DESI BAO+PantheonPlus & $0.303\pm 0.013$ & $9994\pm 110$ & $-0.901\pm 0.057$ & -- \\
\midrule
DESI BAO+Union3 & $0.309\pm 0.013$ & $9835\pm 140$ & $-0.811^{+0.079}_{-0.070}$ & -- \\
\midrule
DESI BAO+DES-SN5YR & $0.306\pm 0.012$ & $9890\pm 100$ & $-0.840\pm 0.052$ & -- \\
\midrule[2pt] 
\textbf{GEDE} & & & & \\
DESI BAO & $0.296^{+0.014}_{-0.016}$ & $10170\pm 330$ & -- & $0.12^{+0.81}_{-1.2}$ \\
\midrule
DESI BAO+PantheonPlus & $0.299\pm 0.013$ & $9993\pm 110$ & -- & $-0.49^{+0.27}_{-0.32}$ \\
\midrule
DESI BAO+Union3 & $0.303\pm 0.014$ & $9845\pm 140$ & -- & $-0.86^{+0.30}_{-0.40}$ \\
\midrule
DESI BAO+DES-SN5YR & $0.302\pm 0.014$ & $9871\pm 99$ & -- & $-0.80^{+0.24}_{-0.27}$ \\
\midrule[2pt] 
\textbf{Mirage} & & & & \\
DESI BAO & $0.430^{+0.066}_{-0.044}$ & $8813^{+300}_{-630}$ & $> 0.078$ & -- \\
\midrule
DESI BAO+PantheonPlus & $0.311\pm 0.012$ & $10010\pm 100$ & $-0.880\pm 0.065$ & -- \\
\midrule
DESI BAO+Union3 & $0.331\pm 0.015$ & $9768\pm 150$ & $-0.67\pm 0.11$ & -- \\
\midrule
DESI BAO+DES-SN5YR &  $0.322\pm 0.011$ & $9875\pm 99$ & $-0.770\pm 0.073$ & -- \\
\midrule
\end{tabular}}
\end{table}
\bibliographystyle{JHEP}
\bibliography{ref,DESI2024_Supporting_refs}
\input{DESI-2024-0420_author_list.affiliations}
\end{document}

%% file: defs.tex
\newcommand{\fde}{\ensuremath{f_{\rm DE}}}

\newcommand{\chisq}{\ensuremath{\chi^2}}

\newcommand{\DVrd}{D_\mathrm{V}/r_\mathrm{d}}
\newcommand{\DMrd}{D_\mathrm{M}/r_\mathrm{d}}
\newcommand{\DHrd}{D_\mathrm{H}/r_\mathrm{d}}

\newcommand{\lcdm}{$\Lambda$CDM}
\newcommand{\wowacdm}{$w_0w_a$CDM}

\newcommand{\Omm}{\ensuremath{\Omega_\text{m}}}
\newcommand{\Omo}{\ensuremath{\Omega_{\text{m},0}}}

\newcommand{\rd}{\ensuremath{r_\text{d}}}

\newcommand{\classy}{\texttt{class}}
\newcommand{\Cobaya}{\texttt{cobaya}}
\newcommand{\gd}{\texttt{GetDist}}
\newcommand{\kmsMpc}{\,{\rm km\,s^{-1}\,Mpc^{-1}}}
\newcommand{\kms}{\,{\rm km\,s^{-1}}}

\crefname{equation}{Eq.}{Eqs.}
\crefname{section}{Section}{Sections}
\crefname{figure}{Fig.}{Figs.}
\crefname{table}{Table}{Tables}
\crefname{appendix}{Appendix}{Appendices}
\Crefname{figure}{Figure}{Figures}
\Crefname{equation}{Equation}{Equations}
\Crefname{section}{Section}{Sections}
\Crefname{table}{Table}{Tables}

\definecolor{llgray}{gray}{0.93}
\definecolor{lgray}{gray}{0.83}
\definecolor{deepmagenta}{rgb}{0.8, 0.0, 0.8}
\definecolor{ballblue}{rgb}{0.13, 0.67, 0.8}
\definecolor{RedWine}{rgb}{0.743,0,0}
\definecolor{celestialblue}{rgb}{0.29, 0.59, 0.82}
\definecolor{DarkGreen}{rgb}{0,0.6,0}



%% file: DESI-2024-0420_author_list.tex
\emailAdd{kushallodha@kasi.re.kr, shafieloo@kasi.re.kr}
\affiliation{Affiliations are in Appendix \ref{sec:affiliations}}

\author[1,2]{{K.~Lodha}\orcidlink{0009-0004-2558-5655},}
\author[1,2]{{A.~Shafieloo}\orcidlink{0000-0001-6815-0337},}
\author[1]{{R.~Calderon}\orcidlink{0000-0002-8215-7292},}
\author[3,4,5]{{E.~Linder}\orcidlink{0000-0001-5536-9241},}
\author[1]{{W.~Sohn},}
\author[6]{{J.~L.~Cervantes-Cota}\orcidlink{0000-0002-3057-6786},}
\author[7]{{A.~de~Mattia},}
\author[8]{{J.~Garc\'ia-Bellido}\orcidlink{0000-0002-9370-8360},}
\author[9]{{M.~Ishak}\orcidlink{0000-0002-6024-466X},}
\author[1]{{W.~Matthewson},}
\author[3]{{J.~Aguilar},}
\author[10]{{S.~Ahlen}\orcidlink{0000-0001-6098-7247},}
\author[11]{{D.~Brooks},}
\author[3]{{T.~Claybaugh},}
\author[12]{{A.~de la Macorra}\orcidlink{0000-0002-1769-1640},}
\author[13]{{A.~Dey}\orcidlink{0000-0002-4928-4003},}
\author[14]{{B.~Dey}\orcidlink{0000-0002-5665-7912},}
\author[11]{{P.~Doel},}
\author[15,16]{{J.~E.~Forero-Romero}\orcidlink{0000-0002-2890-3725},}
\author[17,18,19]{{E.~Gaztañaga},}
\author[3]{{S.~Gontcho A Gontcho}\orcidlink{0000-0003-3142-233X},}
\author[20]{{C.~Howlett}\orcidlink{0000-0002-1081-9410},}
\author[13]{{S.~Juneau},}
\author[21,22]{{S.~Kent}\orcidlink{0000-0003-4207-7420},}
\author[3]{{T.~Kisner}\orcidlink{0000-0003-3510-7134},}
\author[3]{{A.~Kremin}\orcidlink{0000-0001-6356-7424},}
\author[3]{{A.~Lambert},}
\author[3]{{M.~Landriau}\orcidlink{0000-0003-1838-8528},}
\author[23]{{L.~Le~Guillou}\orcidlink{0000-0001-7178-8868},}
\author[24,25,26]{{P.~Martini}\orcidlink{0000-0002-4279-4182},}
\author[13]{{A.~Meisner}\orcidlink{0000-0002-1125-7384},}
\author[27,28]{{R.~Miquel},}
\author[29]{{J.~Moustakas}\orcidlink{0000-0002-2733-4559},}
\author[14]{{J.~ A.~Newman}\orcidlink{0000-0001-8684-2222},}
\author[30,31]{{G.~Niz}\orcidlink{0000-0002-1544-8946},}
\author[7,3]{{N.~Palanque-Delabrouille}\orcidlink{0000-0003-3188-784X},}
\author[32,33,34]{{W.~J.~Percival}\orcidlink{0000-0002-0644-5727},}
\author[3,4,5]{{C.~Poppett},}
\author[35]{{F.~Prada}\orcidlink{0000-0001-7145-8674},}
\author[36]{{G.~Rossi},}
\author[7]{{V.~Ruhlmann-Kleider}\orcidlink{0009-0000-6063-6121},}
\author[37]{{E.~Sanchez}\orcidlink{0000-0002-9646-8198},}
\author[38]{{E.~F.~Schlafly}\orcidlink{0000-0002-3569-7421},}
\author[3]{{D.~Schlegel},}
\author[39,40]{{M.~Schubnell},}
\author[41]{{H.~Seo}\orcidlink{0000-0002-6588-3508},}
\author[13]{{D.~Sprayberry},}
\author[40]{{G.~Tarl\'{e}}\orcidlink{0000-0003-1704-0781},}
\author[13]{{B.~A.~Weaver},}
\author[42]{{H.~Zou}\orcidlink{0000-0002-6684-3997}}

%% file: DESI-2024-0420_author_list.affiliations.tex

\section{Author Affiliations}
\label{sec:affiliations}

\noindent \hangindent=.5cm $^{1}${Korea Astronomy and Space Science Institute, 776, Daedeokdae-ro, Yuseong-gu, Daejeon 34055, Republic of Korea}

\noindent \hangindent=.5cm $^{2}${University of Science and Technology, 217 Gajeong-ro, Yuseong-gu, Daejeon 34113, Republic of Korea}

\noindent \hangindent=.5cm $^{3}${Lawrence Berkeley National Laboratory, 1 Cyclotron Road, Berkeley, CA 94720, USA}

\noindent \hangindent=.5cm $^{4}${Space Sciences Laboratory, University of California, Berkeley, 7 Gauss Way, Berkeley, CA  94720, USA}

\noindent \hangindent=.5cm $^{5}${University of California, Berkeley, 110 Sproul Hall \#5800 Berkeley, CA 94720, USA}

\noindent \hangindent=.5cm $^{6}${Departamento de F\'{i}sica, Instituto Nacional de Investigaciones Nucleares, Carreterra M\'{e}xico-Toluca S/N, La Marquesa,  Ocoyoacac, Edo. de M\'{e}xico C.P. 52750,  M\'{e}xico}

\noindent \hangindent=.5cm $^{7}${IRFU, CEA, Universit\'{e} Paris-Saclay, F-91191 Gif-sur-Yvette, France}

\noindent \hangindent=.5cm $^{8}${Instituto de F\'{\i}sica Te\'{o}rica (IFT) UAM/CSIC, Universidad Aut\'{o}noma de Madrid, Cantoblanco, E-28049, Madrid, Spain}

\noindent \hangindent=.5cm $^{9}${Department of Physics, The University of Texas at Dallas, Richardson, TX 75080, USA}

\noindent \hangindent=.5cm $^{10}${Physics Dept., Boston University, 590 Commonwealth Avenue, Boston, MA 02215, USA}

\noindent \hangindent=.5cm $^{11}${Department of Physics \& Astronomy, University College London, Gower Street, London, WC1E 6BT, UK}

\noindent \hangindent=.5cm $^{12}${Instituto de F\'{\i}sica, Universidad Nacional Aut\'{o}noma de M\'{e}xico,  Cd. de M\'{e}xico  C.P. 04510,  M\'{e}xico}

\noindent \hangindent=.5cm $^{13}${NSF NOIRLab, 950 N. Cherry Ave., Tucson, AZ 85719, USA}

\noindent \hangindent=.5cm $^{14}${Department of Physics \& Astronomy and Pittsburgh Particle Physics, Astrophysics, and Cosmology Center (PITT PACC), University of Pittsburgh, 3941 O'Hara Street, Pittsburgh, PA 15260, USA}

\noindent \hangindent=.5cm $^{15}${Departamento de F\'isica, Universidad de los Andes, Cra. 1 No. 18A-10, Edificio Ip, CP 111711, Bogot\'a, Colombia}

\noindent \hangindent=.5cm $^{16}${Observatorio Astron\'omico, Universidad de los Andes, Cra. 1 No. 18A-10, Edificio H, CP 111711 Bogot\'a, Colombia}

\noindent \hangindent=.5cm $^{17}${Institut d'Estudis Espacials de Catalunya (IEEC), 08034 Barcelona, Spain}

\noindent \hangindent=.5cm $^{18}${Institute of Cosmology and Gravitation, University of Portsmouth, Dennis Sciama Building, Portsmouth, PO1 3FX, UK}

\noindent \hangindent=.5cm $^{19}${Institute of Space Sciences, ICE-CSIC, Campus UAB, Carrer de Can Magrans s/n, 08913 Bellaterra, Barcelona, Spain}

\noindent \hangindent=.5cm $^{20}${School of Mathematics and Physics, University of Queensland, 4072, Australia}

\noindent \hangindent=.5cm $^{21}${Department of Astronomy and Astrophysics, University of Chicago, 5640 South Ellis Avenue, Chicago, IL 60637, USA}

\noindent \hangindent=.5cm $^{22}${Fermi National Accelerator Laboratory, PO Box 500, Batavia, IL 60510, USA}

\noindent \hangindent=.5cm $^{23}${Sorbonne Universit\'{e}, CNRS/IN2P3, Laboratoire de Physique Nucl\'{e}aire et de Hautes Energies (LPNHE), FR-75005 Paris, France}

\noindent \hangindent=.5cm $^{24}${Center for Cosmology and AstroParticle Physics, The Ohio State University, 191 West Woodruff Avenue, Columbus, OH 43210, USA}

\noindent \hangindent=.5cm $^{25}${Department of Astronomy, The Ohio State University, 4055 McPherson Laboratory, 140 W 18th Avenue, Columbus, OH 43210, USA}

\noindent \hangindent=.5cm $^{26}${The Ohio State University, Columbus, 43210 OH, USA}

\noindent \hangindent=.5cm $^{27}${Instituci\'{o} Catalana de Recerca i Estudis Avan\c{c}ats, Passeig de Llu\'{\i}s Companys, 23, 08010 Barcelona, Spain}

\noindent \hangindent=.5cm $^{28}${Institut de F\'{i}sica d’Altes Energies (IFAE), The Barcelona Institute of Science and Technology, Campus UAB, 08193 Bellaterra Barcelona, Spain}

\noindent \hangindent=.5cm $^{29}${Department of Physics and Astronomy, Siena College, 515 Loudon Road, Loudonville, NY 12211, USA}

\noindent \hangindent=.5cm $^{30}${Departamento de F\'{i}sica, Universidad de Guanajuato - DCI, C.P. 37150, Leon, Guanajuato, M\'{e}xico}

\noindent \hangindent=.5cm $^{31}${Instituto Avanzado de Cosmolog\'{\i}a A.~C., San Marcos 11 - Atenas 202. Magdalena Contreras, 10720. Ciudad de M\'{e}xico, M\'{e}xico}

\noindent \hangindent=.5cm $^{32}${Department of Physics and Astronomy, University of Waterloo, 200 University Ave W, Waterloo, ON N2L 3G1, Canada}

\noindent \hangindent=.5cm $^{33}${Perimeter Institute for Theoretical Physics, 31 Caroline St. North, Waterloo, ON N2L 2Y5, Canada}

\noindent \hangindent=.5cm $^{34}${Waterloo Centre for Astrophysics, University of Waterloo, 200 University Ave W, Waterloo, ON N2L 3G1, Canada}

\noindent \hangindent=.5cm $^{35}${Instituto de Astrof\'{i}sica de Andaluc\'{i}a (CSIC), Glorieta de la Astronom\'{i}a, s/n, E-18008 Granada, Spain}

\noindent \hangindent=.5cm $^{36}${Department of Physics and Astronomy, Sejong University, Seoul, 143-747, Korea}

\noindent \hangindent=.5cm $^{37}${CIEMAT, Avenida Complutense 40, E-28040 Madrid, Spain}

\noindent \hangindent=.5cm $^{38}${Space Telescope Science Institute, 3700 San Martin Drive, Baltimore, MD 21218, USA}

\noindent \hangindent=.5cm $^{39}${Department of Physics, University of Michigan, Ann Arbor, MI 48109, USA}

\noindent \hangindent=.5cm $^{40}${University of Michigan, Ann Arbor, MI 48109, USA}

\noindent \hangindent=.5cm $^{41}${Department of Physics \& Astronomy, Ohio University, Athens, OH 45701, USA}

\noindent \hangindent=.5cm $^{42}${National Astronomical Observatories, Chinese Academy of Sciences, A20 Datun Rd., Chaoyang District, Beijing, 100012, P.R. China}